\documentclass[twocolumn,notrackchanges]{aastex63}
\usepackage[english]{babel}
\usepackage{amsmath}

\usepackage{xspace}
\usepackage{units}
\newcommand{\ammo}{\ensuremath{\rm NH_3}\xspace}
\newcommand{\ntwohp}{\ensuremath{\rm N_2H^+}\xspace}
\newcommand{\kms}{\ensuremath{\rm km\,s^{-1}}\xspace}
\newcommand{\kkms}{\ensuremath{\rm K\,km\,s^{-1}}\xspace}
\newcommand{\msun}{\ensuremath{\rm M_\odot}\xspace}
\newcommand{\tk}{\ensuremath{T_{\rm k}}\xspace}
\newcommand{\tex}{\ensuremath{T_{\rm ex}}\xspace}
\newcommand{\sigv}{\ensuremath{\sigma_{\rm v}}\xspace}
\newcommand{\sigth}{\ensuremath{\sigma_{\rm th}}\xspace}
\newcommand{\signt}{\ensuremath{\sigma_{\rm NT}}\xspace}

\newcommand{\vlsr}{\ensuremath{V_{\rm LSR}}\xspace}
\newcommand{\cc}{\ensuremath{\rm cm^{-3}}\xspace}
\newcommand{\ncrit}{\ensuremath{n_{\rm crit}}\xspace}

\shortauthors{J.E. Pineda et al.}
\shorttitle{MHD Wave Penetration}

\received{}
\revised{}
\accepted{}
\submitjournal{ApJ}

\begin{document}

\title{Neutral vs Ion Linewidths in Barnard 5: 
Evidence for Penetration by MHD Waves}

\correspondingauthor{Jaime E. Pineda}
\email{jpineda@mpe.mpg.de}

\author[0000-0002-3972-1978]{Jaime E. Pineda}
\affiliation{Max-Planck-Institut f\"ur extraterrestrische Physik, Giessenbachstrasse 1, 85748 Garching, Germany}

\author[0000-0002-1730-8832]{Anika Schmiedeke}
\affiliation{Max-Planck-Institut f\"ur extraterrestrische Physik, Giessenbachstrasse 1, 85748 Garching, Germany}

\author[0000-0003-1481-7911]{Paola Caselli}
\affiliation{Max-Planck-Institut f\"ur extraterrestrische Physik, Giessenbachstrasse 1, 85748 Garching, Germany}

\author{Steven W. Stahler}
\affiliation{Astronomy Department, University of California, Berkeley, CA 94720, USA}

\author[0000-0003-1924-1122]{David T. Frayer}
\affiliation{Green Bank Observatory, 155 Observatory Road, Green Bank, WV 24944, USA}

\author{Sarah E. Church}
\affiliation{Kavli Institute for Particle Astrophysics and Cosmology; Physics Department, Stanford University, Stanford, CA 94305, USA}

\author[0000-0001-6159-9174]{Andrew I. Harris}
\affiliation{Department of Astronomy, University of Maryland, College Park, MD 20742, USA}

\begin{abstract}
Dense cores are the final place where turbulence is dissipated.
It has been {proposed from theoretical arguments} that the non-thermal velocity dispersion should be narrower both for molecular ions (compared to neutrals) and for transitions with higher critical densities.
To test these hypotheses, we compare the velocity dispersion of \ntwohp(1--0) (\ncrit =  \unit[$6\times10^4$]{\cc}) and \ammo (\ncrit = \unit[$2\times10^3$]{\cc}), in the dense core Barnard 5. 
We analyse well resolved and high signal-to-noise observations 
of \ammo(1,1) and (2,2) obtained with combining GBT and VLA data, and \ntwohp (1--0) obtained with GBT Argus, which present a similar morphology.
Surprisingly, the non-thermal velocity dispersion of the ion is systematically higher than that of the neutral by 20\%.
The derived sonic Mach number, $\mathcal{M}_s = \sigma_{\rm NT}/c_s$, has peak values $\mathcal{M}_{s, \ntwohp} = 0.59$ and $\mathcal{M}_{s, \ammo} = 0.48$ for \ntwohp and \ammo, respectively. 
This observed difference may indicate that the magnetic field even deep within the dense core is still oscillating, as it is in the turbulent region outside the core.
The ions should be more strongly dynamically coupled to this oscillating field than the neutrals, thus accounting for their broader linewidth.
If corroborated by further observations, this finding would shed additional light on the transition to quiescence in dense cores.
\end{abstract}


\section{Introduction} \label{sec:intro}
Molecular cloud  cores, the places where stars form, have been studied using both molecular line transitions of dense gas tracers and dust continuum mapping \citep{BM89,Kauffmann2008-MAMBO}. 
One of the most widely used dense gas tracers is \ammo,  which traces material with densities of a few \unit[$10^3$]{\cc} and higher, and thanks to its hyperfine structure, is useful to determine temperature, 
density, optical depth, and the kinematic properties of the gas \citep{MB83_NH3,BM89}. 
By mapping dense cores in \ammo(1,1), it has been found that they show an 
almost constant level of non--thermal motions, within a certain ``coherence'' zone  
\citep{Goodman_1998-coherence}.
However, the parental molecular cloud hosting the dense core usually displays highly supersonic linewidths \citep{Zuckerman+Evans:1974,Larson_1981-turbulence_MC,2002ApJ...566..289B,Heyer2004-GMC_PCA,Choudhury2020-Letter}.
The term ``coherent core'' is used to describe the dense gas where non--thermal motions 
are roughly constant, and typically smaller than the thermal motions, independent of scale 
\citep[see also][]{Caselli2002_N2Hp_Maps}. 
We have observed the transition from turbulence to quiescence using a single tracer in a few objects at relatively coarse angular resolution  \citep{Pineda:2010jq,Friesen2017}. 
Additional studies at higher resolution, and employing a variety of tracers, should aid our understanding of dense core structure and formation, and hence the initial conditions for star formation 
\citep[e.g.][]{2011ApJ...729..120G,Bailey2015-Transition}.

The empirical linewidth-size relation, first found by \cite{Larson_1981-turbulence_MC} and refined by \cite{Solomon1989}, shows that smaller molecular clouds of higher density are generally more quiescent.
Thus, if this relation applies \emph{within} individual dense cores, we expect that tracers with higher critical density should exhibit narrower line widths \citep{Myers_1983-subsonic_turbulence}. 
So, too, should molecular ion tracers, which are dynamically tethered to the dense core's internal magnetic field.
Neutral species, on the other hand, move about more freely, and are imperfectly coupled to the ions via collisions \citep{Mestel1956-SF_Magnetic_Clouds}.

\cite{Houde2000_Result} compared single dish observations of neutral and ions for which they see a broader line-width for neutrals compared to ions, and derived a relation between the a relation for the line-width of ion and neutral.
Their formalism takes into account the collisions between ion and neutral as particles, under the assumption that ions are forced into gyromagnetic motion about the magnetic field direction, but, it is not a full description of the plasma fluid behavior. 
Under these assumption the ions will always display a narrower linewidth, when tracing the same region.
Recently, MHD simulations show that in some cases the ions could have more energy than neutrals \citep{Burkhart2015-MHD_AD},
however, no clear predictions are made for comparisons with observations inside dense cores  \citep{Meyer2014-Observe_MHD}.
In addition, the observations presented in \cite{Houde2000_Result} are single dish observations towards 
two active star-forming regions, Orion~A and DR21, 
with an angular resolution between 20--30\arcsec, 
which are insufficient to resolve the outflows and substructures observed at higher angular resolution \citep[e.g.][]{Hacar2018-ALMA_OrionA,Monsch2018-VLA_OrionA} and might explain the difference in the HCO$^+$ and HCN lines observed.

Surprisingly, \cite{Tafalla2004-Cores} found that the levels of turbulence displayed by the \ntwohp (1--0) line ($n_{crit}\approx 10^{4}$\,\cc; \citealt{Shirley2015-molecules})
are higher than those obtained with \ammo(1,1)  ($n_{crit}\approx 10^{3}$\,\cc; \citealt{Shirley2015-molecules}) in a few cores observed at coarse angular resolution (40\arcsec) insufficient to resolve the core structure.
Ostensibly, their finding is at odds with the conventional theoretical picture. Until now, however, no data have been of sufficient angular and spectral resolution to 
explore the different kinematics of ion and neutral dense gas tracers on well resolved maps \citep[e.g.][]{Tafalla2004-Cores}.
Therefore, it has been impossible to measure accurately the velocity differences within the subsonic region 
nor the velocity differences between neutral and ions in dense cores.
Similarly, in recent deep observations of \ammo and \ntwohp (1--0) in an IRDC, \cite{Sokolov2019-IRDC_Kinematics} found that \ammo displays non-thermal velocity dispersion $20$\% lower than \ntwohp(1--0). However, those observation lacked sufficient angular resolution to perform a detailed comparison.

Barnard 5 (B5) is an isolated and bright dense core {\citep{Langer1989-B5}} in the Perseus molecular cloud (\unit[302$\pm$21]{pc}, \citealt{Zucker2018b}).  
Its average density in the region of subsonic turbulence is \unit[$\sim7\times 10^{4}$]{\cc}, 
and this region includes a single Class~I object (the \unit[$\sim0.5$]{\msun} B5-IRS1; \citealt{Fuller1991})
{driving a parsec-scale outflow \citep{Langer1989-B5,Bally1996-B5_Outflow,Arce2010-Perseus_Outflows,Zapata2014-B5_Outflow}}.
Previous GBT \ammo(1,1) observations (with 30\arcsec angular resolution) of B5 allowed us to detect for the first time the sharp transition between subsonic and supersonic turbulence in a single tracer \citep{Pineda:2010jq}. 
New \ammo(1,1) observations of B5 obtained with the 
Jansky Very Large Array (JVLA; and combined with the GBT single-dish data) 
showed that the velocity dispersion remains subsonic with little variation even at these high-angular resolutions (6\arcsec). 
These data also revealed the presence of 
filaments within the quiescent dense core,
for which the average radial profile is well fitted by an isothermal cylinder model 
\citep{Ostriker_1964-Filament_Model} and the turbulence is consistently subsonic along them \citep{Pineda2011_B5-VLA}. 
These filaments enclose three high-density bound condensations,  
which combined with the young Class~I object might form a bound quadruple multiple star system \citep{Pineda2015-Multiple}.

We present new observations of the \ntwohp (1--0) line with GBT Argus 
to compare them to those obtained from \ammo (1,1) to measure possible differences between ions and neutrals throughout the dense core. 

\section{Observations}

\begin{figure*}[tbh!]
    \includegraphics[width=\textwidth]{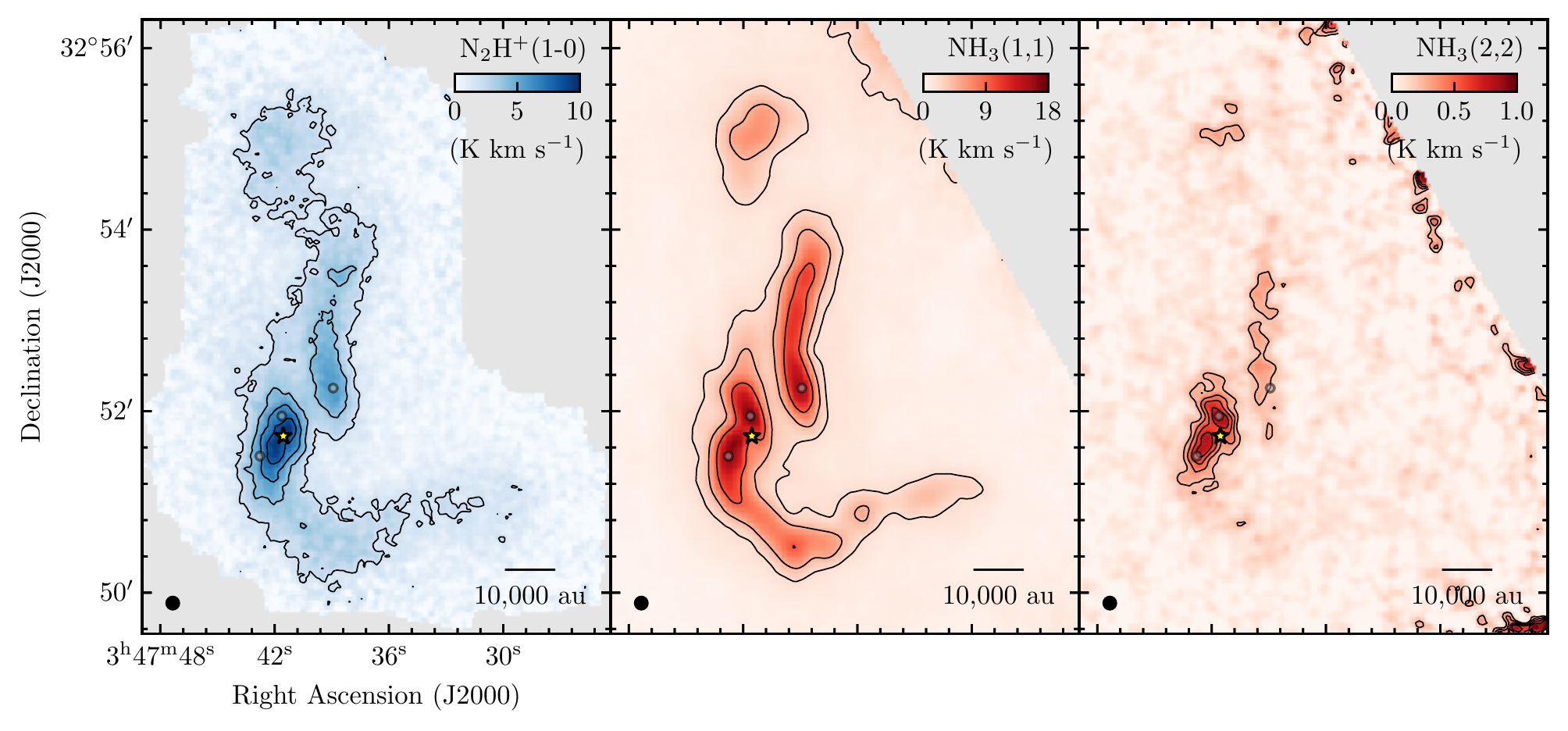}
    \caption{Integrated intensity maps for \ntwohp (1--0), \ammo (1,1), and \ammo (2,2) are shown in left, middle, and right panels, respectively.
    The contour levels are drawn at \unit[[2, 4, 6, 8]]{\kkms}, \unit[[2.4, 4.8, 9.6]]{\kkms}, and \unit[[0.25, 0.4, 0.55, 0.7]]{\kkms} for the \ntwohp, \ammo(1,1), and \ammo(2,2), respectively.
    The star and filled circles mark the positions of the Class I object and the condensations identified by \cite{Pineda2015-Multiple}, respectively.
    The beam and scale bars are shown in the bottom left and right  corners, respectively.
    \label{fig-TdV}}
\end{figure*}

\subsection{\ntwohp}

We observed the \ntwohp (1--0) transition line towards B5 using the Argus receiver at the 100m Robert C. Byrd Green Bank Telescope (GBT) under project GBT/18B-288 on 2018 October 23$^\text{rd}$ and 25$^\text{th}$, and on 2018 November 21$^\text{st}$.
Argus is a 16-pixel W-band array receiver, with a 4$\times$4 configuration \citep{Sieth2014}, operating between 74 and 116\,GHz. 
We observed the blazar 3C84 (J0319+4130), which is an Atacama Large Millimeter/submillimeter Array (ALMA) calibration source \citep{Bonato2018,Bonato2019} to obtain the focus and pointing solutions, and to obtain an absolute calibration of the spectra in units of main beam temperature (T$_\text{MB}$). We configured the VEGAS backend to a rest frequency of \unit[93,173.704]{MHz}, using mode 4 with \unit[187.5]{MHz} of bandwidth and \unit[5.7]{kHz} of spectral resolution. We performed pointing and focus scans every 30--50 min, depending on weather conditions. Science scans were obtained in on-the-fly (OTF) mode and calibration scans were done before and after each block of science scans. The system temperatures are $\approx$118~K for all beams.
The observations were carried out using the frequency switching mode, with a frequency shift of 25\,{MHz} ($\pm$12.5\,{kHz}{MHz} shift). We split B5 into two rectangles (2\arcmin$\times$4\arcmin\,centered on $\alpha_1$=03$^\text{h}$47$^\text{m}$40.78$^\text{s}$,  $\delta_1$=+32$^\circ$53\arcmin43.0\arcsec, and 4\arcmin$\times$2\arcmin\,centered on $\alpha_2$=03$^\text{h}$47$^\text{m}$37.35$^\text{s}$, $\delta_2$=+32$^\circ$50\arcmin43.6\arcsec) and scanned both sub-maps repeatedly in orthogonal scanning directions (i.e. along right ascension and declination) to reduce the effect of striping and minimize atmospheric instabilities affecting the data quality.

Standard calibration was performed using the GBTIDL package \citep{Marganian2006_GBT} following the procedures and algorithms appropriate for Argus \citep{Frayer2019-Argus_Calibration}.
The calibrated data was mapped using the \verb+griddata+ task from the python-based package \verb+gbtpipe+\footnote{\url{https://github.com/GBTSpectroscopy/gbtpipe}}. 
At that stage of data calibration we also performed the baseline subtraction, by fitting a polynomial (blorder=5) to line-free channels before gridding the data.
Using a main beam efficiency of 0.46$\pm$0.07, we converted from $T_A^*$ to $T_\text{MB}$. 
Then we perform a final first order polynomial baseline removal using the GAS pipeline \citep{Friesen2017}. 
We smooth the data cube spatially to a final beam size of \unit[8]{\arcsec} and spectrally to a final spectral resolution of \unit[0.049]{\kms} using the \texttt{SpectralCube} Python package. 
The mean rms in the 4\arcmin$\times$4\arcmin\,map is \unit[290]{mK}.
The integrated intensity map is calculated over all the hyperfine components and shown in the left panel of Fig.~\ref{fig-TdV}.
The noise level of the integrated intensity is estimated as \unit[0.4]{\kkms} using the emission free region of the map.

\subsection{\ammo}
We use the \ammo (1,1) and (2,2) data from Karl G. Jansky Very Large Array (VLA) and Green Bank Telescope (GBT) presented in \cite{Pineda2015-Multiple}.
The data are a 27-pointing mosaic with the VLA, which is then imaged using multiscale clean and combined with the single dish data (GBT) using the model image. 
We finally smooth the data cube to the same angular resolution and re-grid the cube to match the \ntwohp(1--0) Argus data (\unit[8]{\arcsec}) using the \texttt{SpectralCube} and  \texttt{reproject} Python packages.
The typical noise level is \unit[100]{mK} and \unit[150]{mK} for the \ammo(1,1) and (2,2), respectively.
The spectral resolution is \unit[0.049]{\kms} for both cubes.
The integrated intensity maps, calculated over all the hyperfine components, are shown in the middle and right panel of Fig.~\ref{fig-TdV}.
The noise level of the integrated intensity are estimated as \unit[0.3]{\kkms} and \unit[0.05]{\kkms} using the emission free region of the map for the \ammo(1,1) and \ammo(2,2) maps, respectively.

\section{Results}
\subsection{Line fit}
We perform the line fit for both species, \ntwohp and \ammo, using the \verb+n2hp+ and \verb+cold-ammonia+ models implemented in \verb+pyspeckit+ \citep{pyspeckit}.

The \ntwohp (1--0) is modeled using the hyperfine structure and relative weights from \cite{Pagani2009-frequencies}
and L. Dore (2011, private communication). 
The model assumes a Gaussian velocity distribution (with central velocity and velocity dispersion \vlsr and \sigv, respectively), equal excitation temperatures (\tex) for all hyperfine components, and total optical depth ($\tau_0$).
We perform an ``optically thick'' fit for the whole cube with all four free parameters, and an ``optically thin'' fit with $\tau_0 = 0.1$ as a fixed parameter. 
We use the results of the optically thick fits, except for pixels where the optical depth has a signal-to-noise less than three ($\tau_0 < 3\sigma(\tau_0)$) for which we use the optically thin fit.
We discard all velocity determinations if the uncertainty in the \sigv is larger than \unit[0.02]{\kms}.

We simultaneously fit the \ammo(1,1) and (2,2) line profiles using the \verb+cold-ammonia+ model \citep{Friesen2017}, which gives a centroid velocity (\vlsr), velocity dispersion (\sigv), kinetic temperature (\tk), excitation temperature (\tex), and total column density of \ammo ($N(\ammo)$, where an ortho-to-para ratio of 1 is assumed). This model assumes that, because of the low kinetic temperatures, only levels (1,1) and (2,2) are populated.
The mean value of the kinetic temperature (\tk) in pixels with uncertainty better than \unit[1]{K} is \unit[9.7]{K}.
For pixels with uncertainties in \tk larger than \unit[1]{K} we re-run the fit, but with a fixed \tk of \unit[9.7]{K}.
The kinematic parameters, \vlsr and \sigv, are well determined even in cases where the kinetic and excitation temperatures are poorly constrained, thanks to the many hyperfine components.
Similarly to \ntwohp, we discard all velocity determinations using \ammo if the uncertainty in the \sigv is larger than \unit[0.02]{\kms}.
We consider that the \tex and \tk are well determined only when their derived uncertainties are smaller than \unit[1]{K}, 
and that the column density is well constrained where \tex and \tk are well constrained.

The velocity dispersion derived from both species are further corrected by the channel response:
\begin{equation}
    \sigv^2 = \sigma_{\rm v,fit}^2 - 
              \left( \frac{\Delta_{\rm ch}}{2.355}\right)^2~,
\end{equation}
where $\sigma_{\rm v,fit}$ is the resulting velocity dispersion from the fit, and $\Delta_{\rm ch}$ is the channel width from the observations, 0.049 \kms for \ntwohp and \ammo.

Some representative spectra are shown in Appendix~\ref{Appendix-spectra}, with the best fit models.
There we show that the models are a good fit, and that there are no strong non-LTE effects on the \ntwohp(1--0) line which could affect the reliability of the derived kinematical parameters.

\subsection{Comparison between tracers}
\begin{figure*}[htb!]
\centering
    \includegraphics[width=\textwidth]{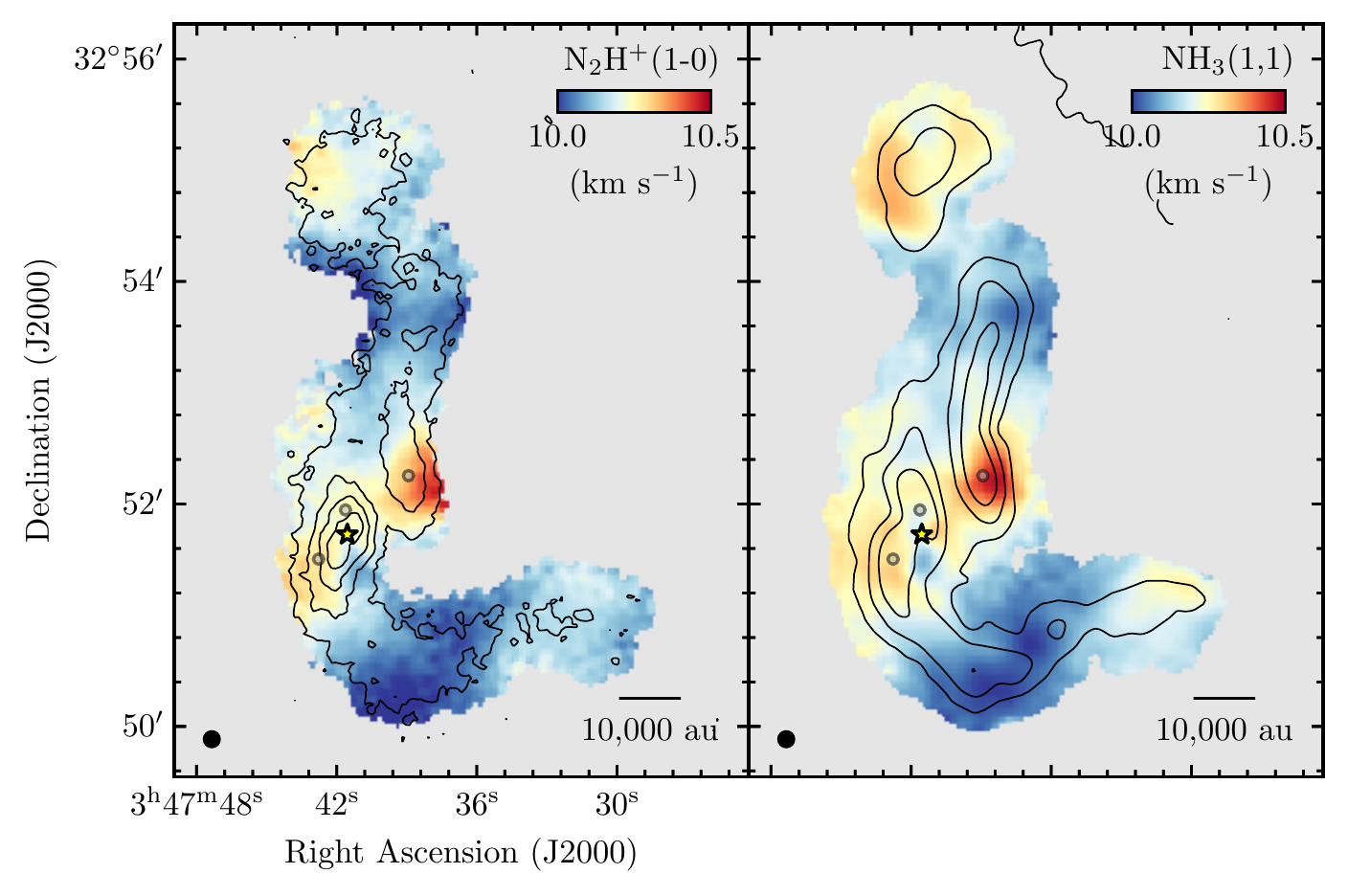}
    \caption{Centroid velocity maps. Left and right panels show the results for \ntwohp and \ammo, respectively.
    The contours are shown for the \ntwohp and \ammo(1,1) integrated intensity maps for the left and right panel, respectively, with levels as in Figure~\ref{fig-TdV}. 
    The star and gray circles mark the positions of the Class I object and the condensations identified by \cite{Pineda2015-Multiple}, respectively.
    The beam and scale bars are shown in the bottom left and right  corners, respectively.
    \label{fig:vlsr}}
\end{figure*}

\begin{figure*}[tb!]
\centering
    \includegraphics[width=0.5\textwidth]{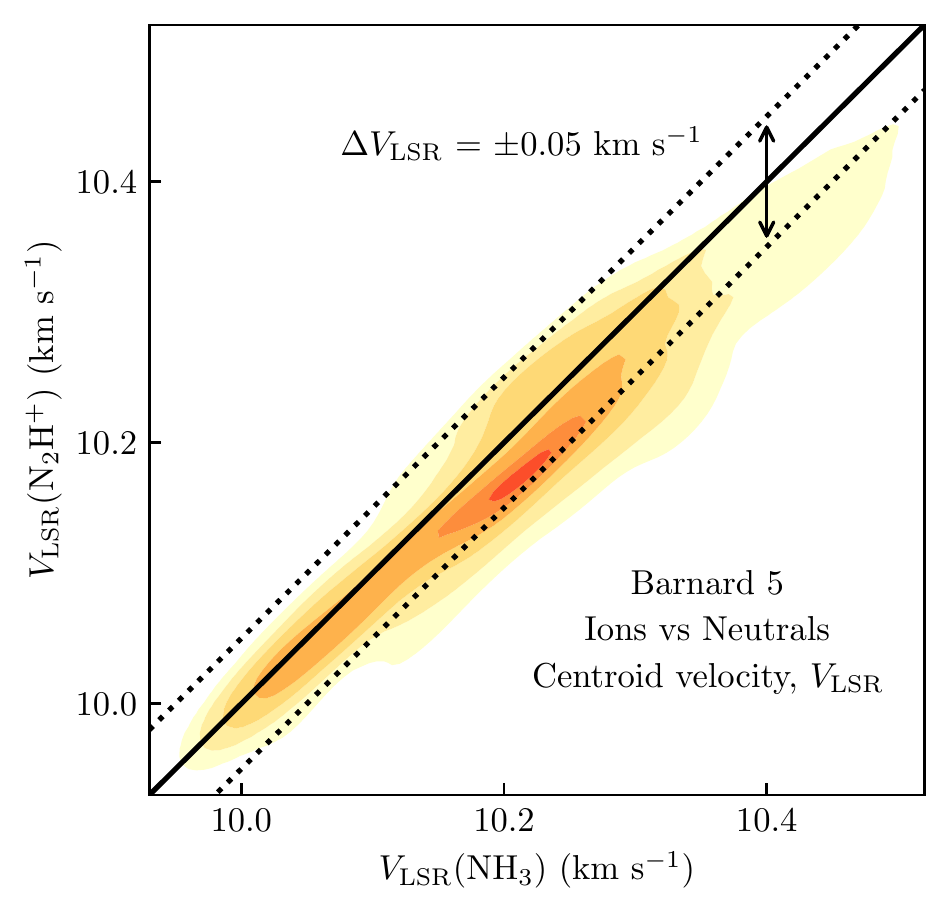}
    \includegraphics[width=0.45\textwidth]{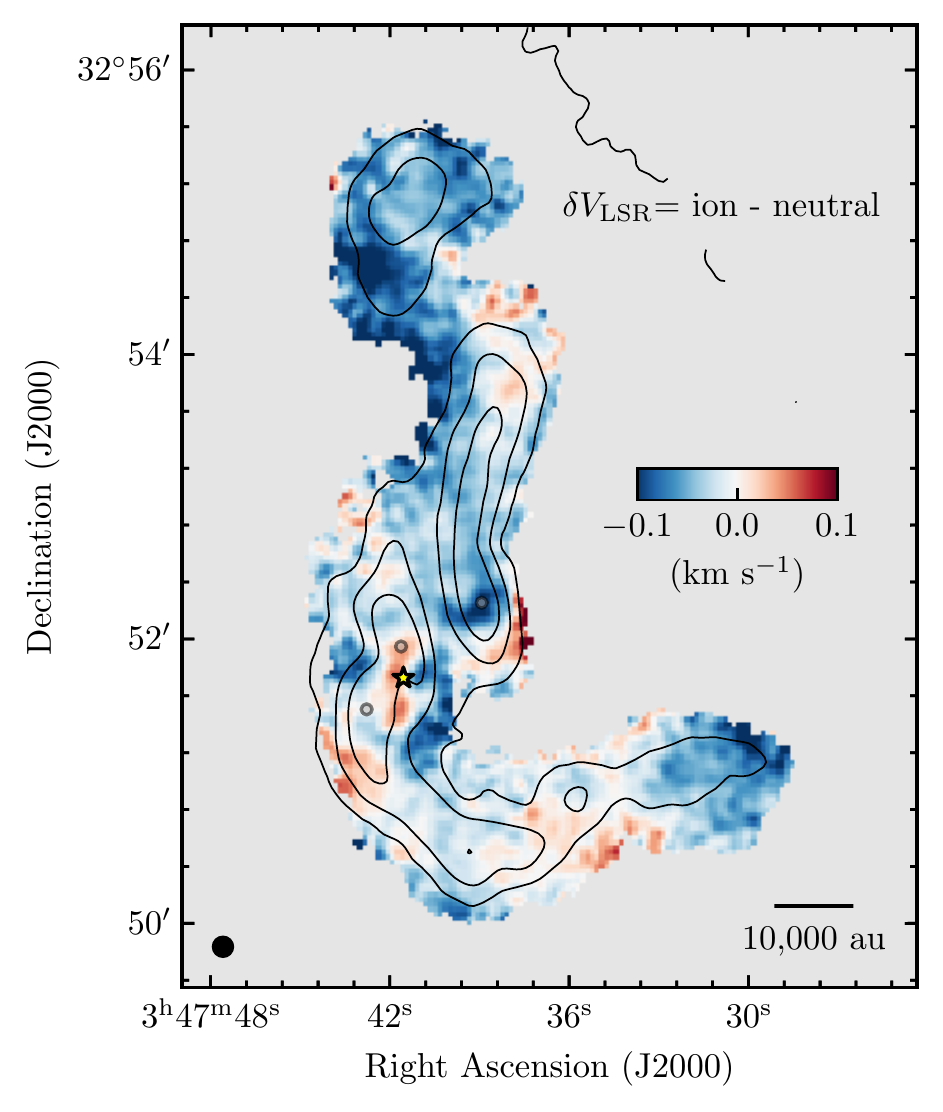}
    \caption{{\it Left:} Comparison between centroid velocity of \ntwohp and \ammo, estimated using a two-dimensional kernel density estimation (KDE).
    The one-to-one line is marked by the black solid line, as well as off-sets of \unit[$\pm$0.05]{\kms} with the dotted lines. 
    {\it Right:} Spatial map showing the difference between ions and neutrals centroid velocity (\vlsr(\ntwohp)$-$\vlsr(\ammo)).
    Blue hues show that ions present a smaller centroid velocity than neutrals. 
    The contours are shown for the \ammo (1,1) integrated intensity map, with levels as in Figure~\ref{fig-TdV}.
    The star and gray circles mark the positions of the Class I object and the condensations identified by \cite{Pineda2015-Multiple}, respectively.
    The beam and scale bar are shown in the bottom left and right corners, respectively.
    \label{fig:compare_vlsr_kde}}
\end{figure*}

The centroid velocity maps for both tracers are shown in Figure~\ref{fig:vlsr}.
The same color scale and stretch is used between left and right panels to allow a direct side-by-side comparison.
The centroid velocity maps for both tracers show a similar pattern, and it suggests that both trace similar material. 
The left panel of Figure~\ref{fig:compare_vlsr_kde} shows the comparison of the centroid velocity obtained for both tracers using kernel density estimation  \citep[KDE;][]{2020SciPy-NMeth}, which agrees within \unit[0.05]{\kms}. 
For the KDE determination we use a relative weight for each data point of $(\sigma(\vlsr(\ammo))\cdot\sigma(\vlsr(\ntwohp)))^{-1}$.
This comparison shows that there is not a single velocity offset between the different tracers. 
The right panel of Figure~\ref{fig:compare_vlsr_kde} shows the difference in velocity map between the ion and neutral. 
This map shows that neutrals display mostly redder velocities than ions, but that in a few regions the relation is the reverse, although without a clear spatial trend.

\begin{figure*}[hbt!]
\centering
    \includegraphics[width=\textwidth]{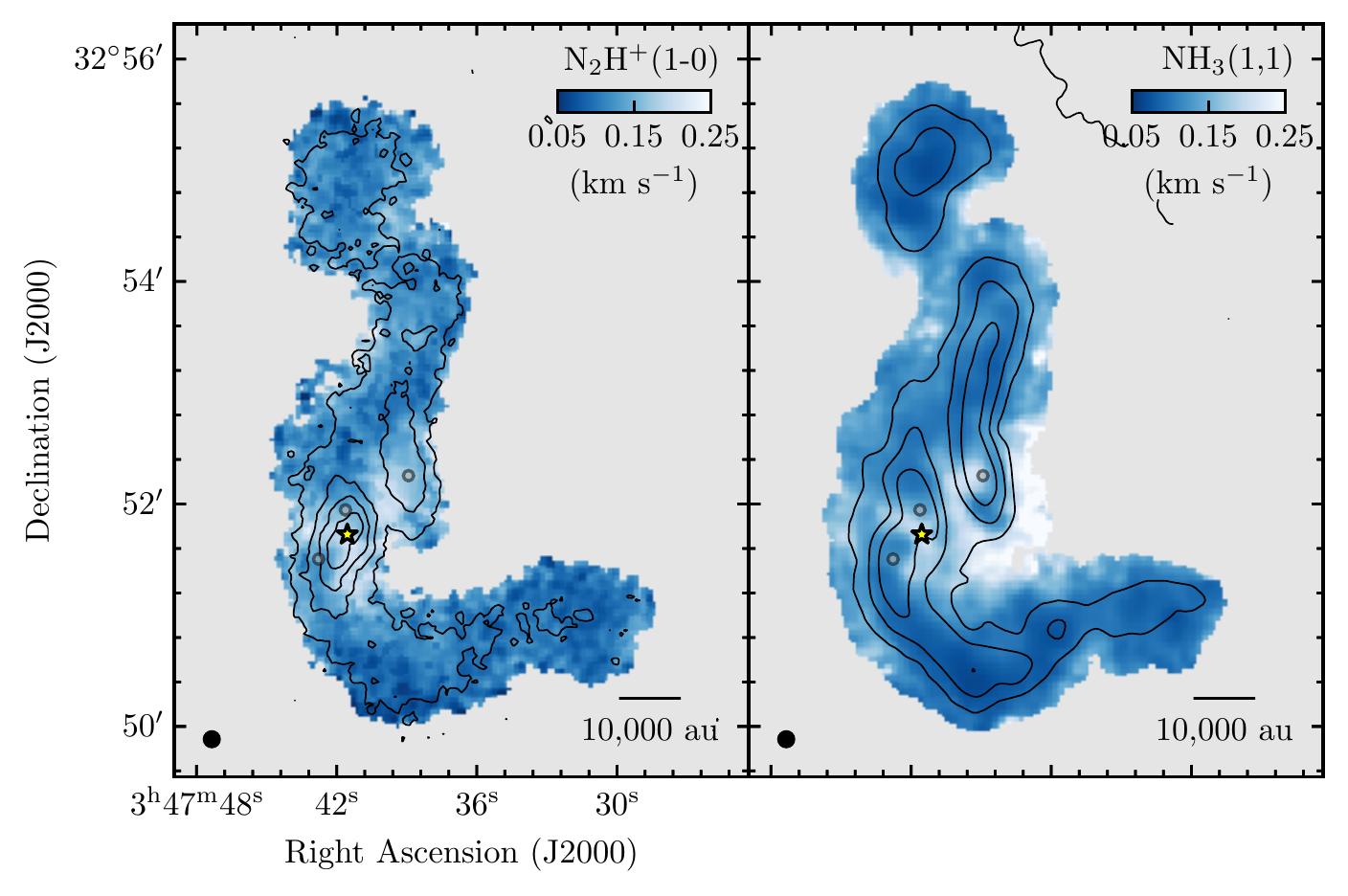}
    \caption{
    Velocity dispersion maps for \ntwohp and \ammo are shown in the left and right panels, respectively.
    The contours are shown for the \ntwohp (1--0) and \ammo(1,1) integrated intensity maps for the left and right panel, respectively, with levels as in Figure~\ref{fig-TdV}. 
    The star and gray circles mark the positions of the Class I object and the condensations identified by \cite{Pineda2015-Multiple}, respectively.
    The beam and scale bars are shown in the bottom left and right  corners, respectively.
    \label{fig:sigma_v}}
\end{figure*}

The velocity dispersion map for both tracers are shown in Figure~\ref{fig:sigma_v}.
The same color scale and stretch is used between left and right panels to allow a direct side-by-side comparison.
The velocity dispersion maps for the two tracers also show a similar pattern, as seen in Fig.~\ref{fig:sigma_v}. 
In this figure, the two distributions of velocity dispersion look quite similar.

\begin{figure}[tb!]
\centering
    \includegraphics[width=\linewidth]{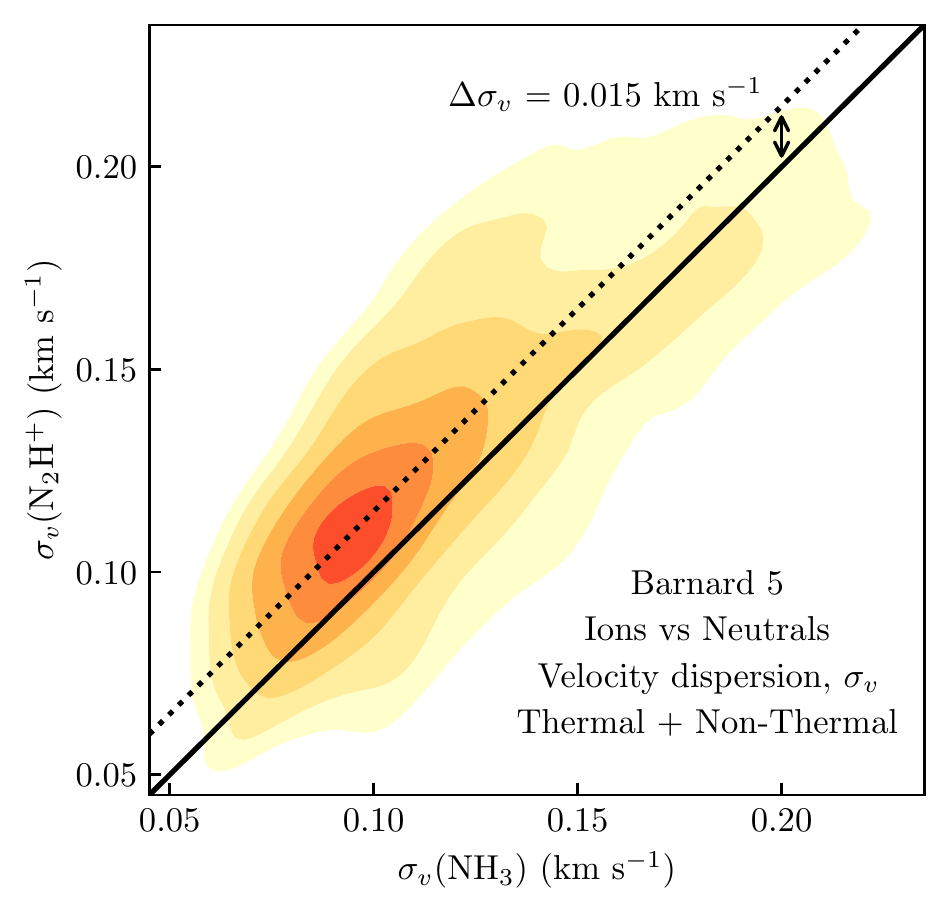}
    \caption{Comparison between velocity dispersion derived from \ntwohp (1--0) and \ammo (1,1), estimated using a two-dimensional KDE. 
    It clearly shows that the \ntwohp (higher density tracer and ion) has a systematically broader velocity dispersion than \ammo (lower-density tracer than \ntwohp and a neutral).
    The one-to-one line is marked by the black solid line, while the dotted line marks $\sigv(\ntwohp) = \sigv(\ammo)+$\unit[0.015]{\kms}.
    \label{fig:compare_sigmav_kde}}
\end{figure}

\begin{figure}[tb!]
\centering
    \includegraphics[width=\linewidth]{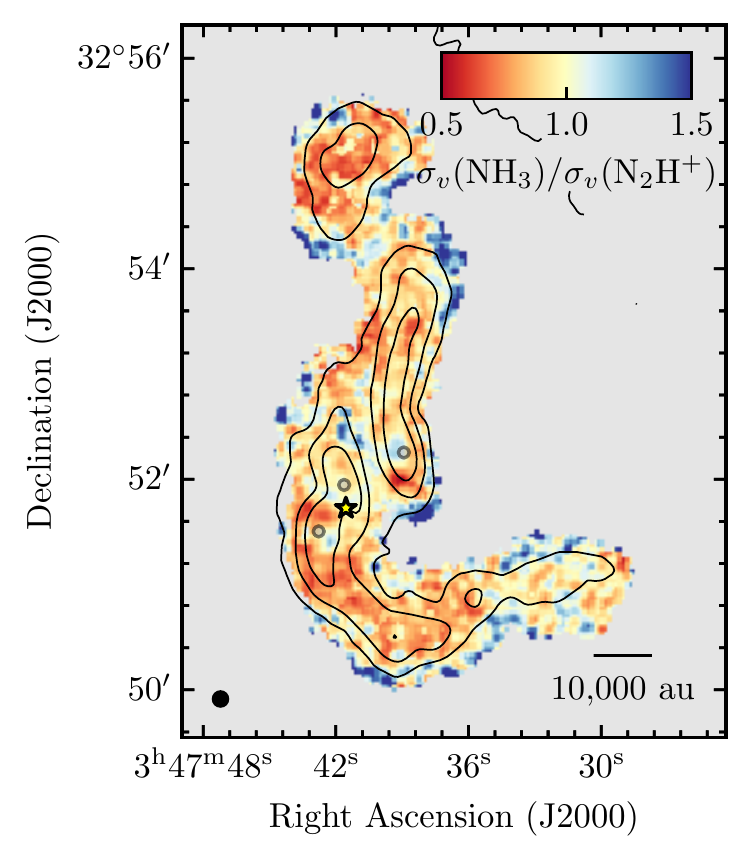}
    \caption{
    Ratio of the velocity dispersion between \ammo (1,1) and \ntwohp (1--0).
    The prevalence of red hues shows that \ammo displays lower velocity dispersion across most of the core than \ntwohp.
    The contours are shown for the \ammo (1,1) integrated intensity map, with levels as in Figure~\ref{fig-TdV}.
    The star and gray circles mark the positions of the Class I object and the condensations identified by \cite{Pineda2015-Multiple}, respectively.
    The beam and scale bars are shown in the bottom left and right corners, respectively.
    \label{fig:ratio_sigmav}}
\end{figure}

However, a direct comparison of the velocity dispersion distributions for all pixels, see Figure~\ref{fig:compare_sigmav_kde}, shows that \ammo velocity dispersions are systematically narrower than those derived using \ntwohp by \unit[$\approx 0.015$]{\kms}.
We calculate the ratio between the velocity dispersion obtained between \ntwohp and \ammo ($R(\ammo/\ntwohp)$), see Figure~\ref{fig:ratio_sigmav}, which shows that the velocity dispersion of \ntwohp is systematically larger throughout the area observed.
There is not clear regional or morphological trend, except that the only places where the velocity dispersion of \ammo is broader than of \ntwohp are close the edge of the map.
It is possible that since \ammo (1,1) can also be excited at densities around \unit[$10^3$]{\cc} and our observations are deep, then these \ammo observations are capable of detecting the brighter pixels of the broader \ammo lines seen during the transition to coherence, previously detected at coarser resolution  \citep{Pineda:2010jq,Friesen2017,Choudhury2020-Letter}.

\subsection{Non-Thermal velocity dispersion comparison}
Thanks to the well constrained kinetic temperature, \tk, from the \ammo observations, we estimate the thermal component of the velocity dispersion, \sigth, to derive a non-thermal velocity dispersion, \signt, as 
\begin{equation}
    \signt^2 = \sigv^2 - \sigth(\tk)^2~, \label{eq:sigma_nt}
\end{equation}
where \sigth is the thermal velocity dispersion of the observed species,
\begin{equation}
    \sigth = \sqrt{\frac{k_{\rm B} \tk}{\mu\,m_{\rm H}}}~, \label{eq:sigma_th}
\end{equation}
where $\mu$ is the molecular weight of the species studied, $m_{\rm H}$ is the hydrogen mass, and $k_{\rm B}$ is the Boltzmann's constant.

\begin{figure}[htb!]
\centering
    \includegraphics[width=\linewidth]{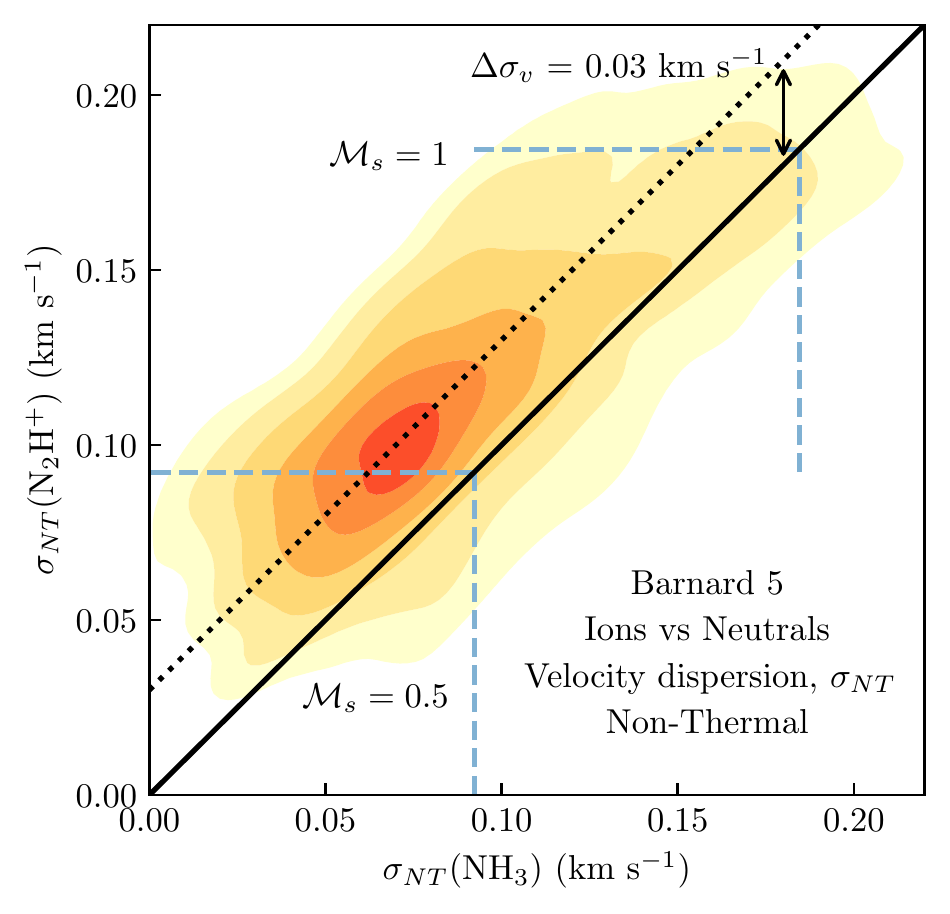}
    \caption{Comparison of non-thermal velocity dispersion derived from \ntwohp (1--0) and \ammo (1,1), estimated using a two-dimensional KDE. 
    It clearly shows that the \ntwohp (higher density tracer and ion) has a systematically broader velocity dispersion than \ammo (lower-density tracer than \ntwohp and a neutral).
    The one-to-one line is marked by the black solid line, while the dotted line marks $\sigv(\ntwohp) = \sigv(\ammo)+$\unit[0.03]{\kms}.
    The light blue lines mark the sonic Mach number values of 0.5 and 1 for the gas at \unit[9.7]{K} and a mean molecular weight of 2.37$\times m_{\rm H}$.
    \label{fig:compare_sigmav_nt_kde}}
\end{figure}

The comparison between the different species is shown in Figure~\ref{fig:compare_sigmav_nt_kde}, which shows that \ntwohp displays systematically and significantly higher levels of the non-thermal component, about \unit[0.03]{\kms}, than \ammo.

We determine the level of non-thermal velocity dispersion (\signt) for both species using the kinetic temperature measured from \ammo and equation~\ref{eq:sigma_nt}.
The sonic Mach number of the turbulence is estimated as
\begin{equation}
    \mathcal{M}_{s} = \frac{\signt}{c_s}~,
\end{equation}
where $c_s$ is the sound speed of the mean particle (using equation~\ref{eq:sigma_th} with $\mu=2.37$, \citealt{Kauffmann2008-MAMBO}).

\begin{figure}[htb!]
\centering
    \includegraphics[width=\linewidth]{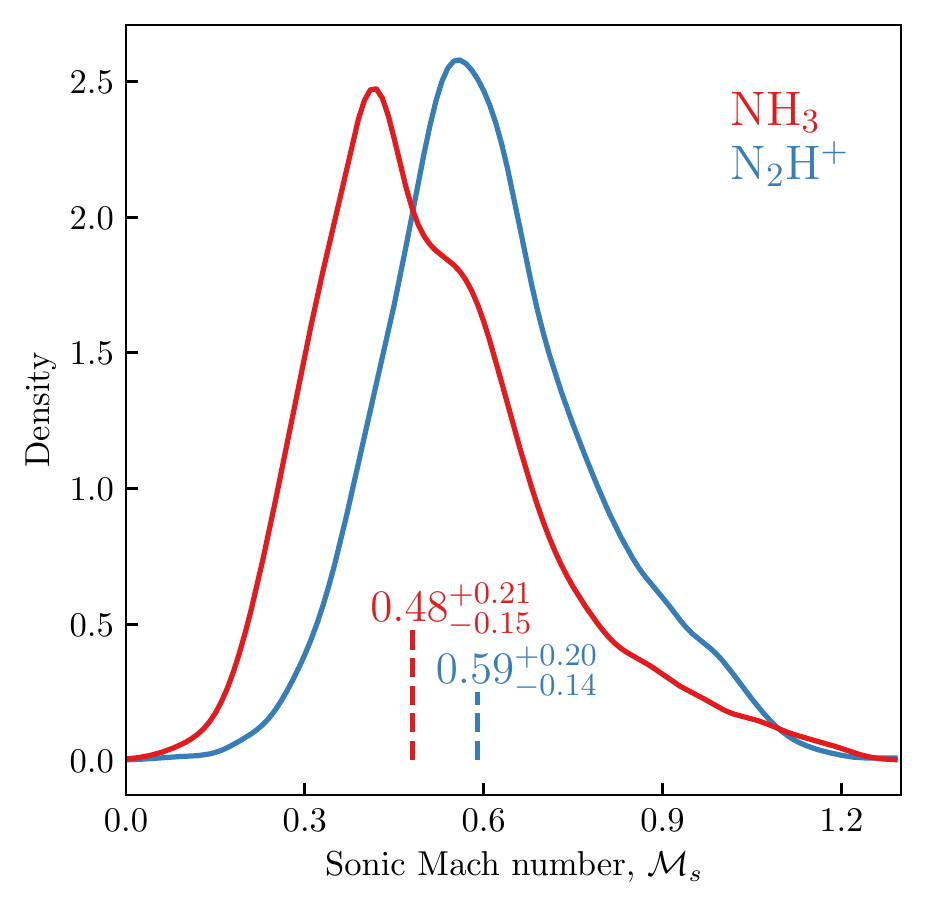}
    \caption{Sonic Mach number distributions for \ntwohp and \ammo, estimated using KDE, are shown in blue and red, respectively. 
    The median values of each sample are marked by the vertical dashed lines. 
    It clearly shows that the \ntwohp (higher density tracer and ion) displays a larger sonic Mach number than \ammo (lower-density tracer than \ntwohp and a neutral).
    The sonic Mach number are derived using the \tk as the gas temperature and a mean molecular weight of of 2.37$\times m_H$.
    \label{fig:compare_kde_Mach_s}}
\end{figure}

Figure~\ref{fig:compare_kde_Mach_s} shows the distribution of Mach numbers for the two tracers using KDE. 
The \ammo distribution, shown in blue, displays a Mach number much smaller (median value of 0.48) than the one in \ntwohp (median value of 0.59).
This substantial difference in the Mach number is present independently on the actual kinetic temperature used, because already the velocity dispersion comparison shows the same trend, while the thermal component of \ammo should be larger than that of \ntwohp because of the smaller in molecular weight.

The observations analysed already resolve the substructure found in the coherent region, the filaments and fragments.
If the difference in observed velocity dispersion were due to still unresolved substructure with a more complex gas dynamics, then this disagreement would be lower in regions away from the higher density filaments. 
However, the difference is also seen in the surrounding material  and lower density regions of the filament.
This suggests that the difference is not due to unresolved structures in the region.

\section{Discussion}
The comparison of both the morphology of the integrated intensity and the similar pattern seen in the centroid velocity (see Figures~\ref{fig-TdV} and \ref{fig:vlsr}) suggests that \ammo and \ntwohp trace similar structures and therefore the direct comparison is fair.

\subsection{Comparison to Previous Predictions}
The critical densities for \ammo (1,1) and \ntwohp (1--0) at \unit[10]{K} are \unit[$2\times10^3$]{\cc} and \unit[$6\times10^4$]{\cc}, respectively \citep[see][]{Shirley2015-molecules}.
If these transitions are tracing material close to these critical densities, and if higher-density gas is concentrated in smaller volumes,
then \ammo {should} trace larger scales than \ntwohp.
Indeed, our observations of B5 presented in Fig.~\ref{fig-TdV} confirm that in B5 the neutrals (\ammo) trace scales larger than the ions (\ntwohp) \citep[see also][]{Pineda:2010jq}.
{Finally, simulations of the} turbulent cascade  \citep{Ballesteros-Paredes2007-PPV_Turbulence} implies that the transition tracing 
smaller scales (the higher density tracer) does present the smaller Mach Number.

Similarly, the presence of static magnetic fields in this star-forming region would give the ions, which are tethered to the field, a smaller velocity dispersion than the neutrals \citep{Li_2008-Turbulence_Dissipation}. 
For the tracers used in this work the predictions from \cite{Houde2000_Result} and  \cite{Li_2008-Turbulence_Dissipation} are opposite to our observations.

\subsection{Magnetic Support of Barnard 5}
Despite the large mass seen in the coherent core, there is no evidence of large scale infall motions  \citep{Campbell2016-Infall_Search}, suggesting that the whole region is not undergoing a dynamical collapse.
However, there is not sufficient support against gravity by turbulence, therefore other components might help to provide support to this region.

By analyzing more carefully the structure of the filaments, we have estimated an upper limit to the magnitude of the {\it static} magnetic field required to support them as \unit[$B_o\approx$500]{$\mu$G}  \citep{Schmiedeke2021-B5} using the relation from \cite{Fiege2000}, this magnetic field strength is larger than those estimated in other starless cores \citep[$\sim$200 $\mu$G,][]{Liu2019-JCMT-POL2_OphC,Crutcher2004-SCUBA}.

\subsection{Penetration by MHD Waves}
The transition to coherence that demarcates the outer boundary of dense cores is only partially understood, and remains a topic of ongoing investigation. 
Several recent studies, also observing the \ammo(1,1) line, found this transition to be remarkably sharp, less than \unit[0.04]{pc} in spatial extent 
\citep{Pineda:2010jq,Friesen2017,Chen2019-Droplets,Auddy2019-Bfield}
In particular, the sudden decrease in velocity dispersion seen in the neutrals may not apply to the ions.

Ever since the discovery of supersonic motions in the larger cloud medium, it has been {proposed} that these supersonic linewidths indicates the presence of MHD waves \citep{Arons1975}.
Of the possible modes, most power is probably contained in Alfv\'en waves, which are incompressible and therefore less subject to dissipation \citep{Zweibel1983}. 
Of course, it is only the charged species that respond directly to the fluctuating magnetic field. 
Neutrals are coupled via collisions to the ions. 
In a static, or slowly changing, field, the two species undergo the relative slip that constitutes ambipolar diffusion \citep{Mestel1956-SF_Magnetic_Clouds}. 
The same slippage occurs in rapidly fluctuating MHD waves, as was shown in the classic study of \cite{Kulsrud1969}, see also \citep{Mouschovias2011-MHD} for a more recent overview.

Our confirmation of enhanced linewidth for ions with respect to neutrals in the deep interior of B5 suggests a modification to the picture for the transition to coherence. Entering the core from the outside, the steep rise in ambient density leads to a falloff in ion fraction, and therefore in dynamical coupling between ions and neutrals. 
In particular, the sudden decrease in neutral velocity may not be present in either the ions or the MHD waves impinging on the core. 

Both species in our study have subsonic non-thermal velocity dispersions. Thus, the vigorous waves permeating the external medium are damped further inside. This damping is caused by ion-neutral friction. The waves persist into the deep interior, but at reduced amplitude. At the core boundary, where neutral velocities are sharply reduced, the waves fluctuations undergo a more gradual transition. As always, the ions track the wave motion, and thus have higher velocity dispersion. If this picture is correct, it should be possible to corroborate with additional observations of both charged and neutral species.  

\subsubsection{Wave Amplitude}
In the case where ions and neutrals are not perfectly coupled, then we can relate the ion and neutral velocities using equation 10.19 from \cite{StahlerPalla-2004}
\begin{equation}
\delta u = \delta u_i \left(1 - \frac{i\omega}{n_i\langle \sigma_{in} u_i' \rangle}\right)^{-1}~,
\label{eq:1}
\end{equation}
where 
$\omega$ is the wave frequency, 
$n_i\langle \sigma_{in} u_i' \rangle$
is the frequency with which a given natural atom or molecule is struck by ions, $\delta u$ is the perturbation on the neutral's velocity, and $\delta u_i$ is the perturbation on the ion's velocity.
Moreover, in this region $\omega$ and the wavenumber, $k$, are related  
in the long wavelength limit \citep{Pinto2012-MHD_Waves} 
as
\begin{equation}
\frac{\omega^2}{k^2} = \frac{B_o^2}{4\pi \rho_o} 
                       \left(1 - \frac{i\omega}{n_i\langle \sigma_{in} u_i' \rangle}\right)~,
\label{eq:2}
\end{equation}
which is equation 10.21 from \cite{StahlerPalla-2004}, 
where $B_o$ is the unperturbed magnetic field and 
$\rho_o$ is the gas density.
Finally, the wave dispersion relation \citep[equation 10.17 from][]{StahlerPalla-2004} relates the perturbation in the magnetic field, $\delta B$, as
\begin{equation}
\delta u_i = -\frac{\omega}{k} \frac{\delta B}{B_o}~.
\label{eq:3}
\end{equation}
{Note that in equations \ref{eq:1}, \ref{eq:2} and \ref{eq:3}, $\omega$ is complex and can be written as  $\omega=\omega_R + i\,\omega_I$.}

We estimate the amplitude of the velocity perturbations as the velocity dispersions, 
{$\delta u =\sqrt{2}\, \sigma_{\rm NT}$,} derived from the spectral lines\footnote{This relates the dispersion and amplitude of a sinusoidal wave.} and therefore rewrite equation \ref{eq:1} as 
\begin{equation}
\left| \frac{\delta u_i}{\delta u} \right|^2 =
1 + 
\frac{2\,\omega_I}{\omega_o}
+
\frac{|\omega|^2}{\omega_o^2}
\approx \left(\frac{\signt({\rm N_2H^+})}{\signt({\rm NH_3})}\right)^2,
\label{eq:4}
\end{equation}
where $\omega_I$ is the imaginary component, 
and we define 
$\omega_o \equiv n_i\langle \sigma_{in} u_i' \rangle$.

Now, equation \ref{eq:2} can also be rewritten as
\begin{eqnarray}
\left|\frac{\omega^2}{k^2}\right| &=& \left(\frac{B_o^2}{4\pi \rho_o}\right)
                       \left[1 + 
\frac{2\,\omega_I}{\omega_o}
+
\frac{|\omega|^2}{\omega_o^2}\right]^{1/2} \nonumber\\
&\approx& \frac{B_o^2}{4\pi \rho_o}
\frac{\signt({\rm N_2H^+})}{\signt({\rm NH_3})}~,
\label{eq:5}
\end{eqnarray}
which relates $\omega$ and $k$ with observables quantities.

Combining equations \ref{eq:3}, \ref{eq:4} and \ref{eq:5} we obtain
\begin{equation}
\frac{\delta B}{B_o} = \sqrt{\signt({\rm N_2H^+})\, \signt({\rm NH_3})}
\frac{\sqrt{8\pi \rho_o}}{B_o}
\end{equation}
or
\begin{equation}
\delta B= \sqrt{8\pi \rho_o} \sqrt{\signt({\rm N_2H^+})\, \signt({\rm NH_3})}~,
\label{eq:deltaB}
\end{equation}
which reveals that the magnetic field perturbation is independent on the mean magnetic field strength, but dependent on the density of the gas traced and the geometrical mean of the ions and neutral non-thermal velocity dispersions observed.

The term $n_i$ is estimated using the ionization degree,
\begin{equation}
x(e) = \frac{n_i}{n({\rm H_2})} = 5.2\times 10^{-6} \left( \frac{n({\rm H_2})}{{\rm cm}^{-3}}\right)^{-0.56} ~,
\end{equation}
from \cite{Caselli2002_L1544-ions}, which is appropriate for dense cores where there is depletion.
The term relating the ions and neutral collisions is approximated by the Langevin term,
\begin{equation}
\langle \sigma_{in} u_i' \rangle = 1.69\times 10^{-9}\, {\rm cm^3\,s^{-1}}~,
\end{equation}
for HCO$^+$--H$_2$  collisions 
\citep{McDaniel:1973vs}.

For the values obtained inside the coherent core, median and 1-sigma spreads of the distributions 
$\langle\signt(\ntwohp)\rangle = 0.109_{-0.027}^{+0.038}$ \kms and 
$\langle\signt(\ammo)\rangle = 0.088_{-0.028}^{+0.040}$ \kms, we estimate the 
\begin{equation}
    \delta B_{in} \approx 27\, \mu{\rm G}~,
\end{equation}
which is at least 5\% of the magnetic field strength estimated inside the coherent core.

\subsubsection{Estimate of Wavelength}
Another important parameter to constrain is the wave's wavelength 
\begin{equation}
\lambda = \frac{2\pi}{\omega} \frac{\omega}{k}~.
\end{equation}
The frequency inside the core, $\omega_{in}$, can be estimated, if we assume that the imaginary component $\omega_I$ is relatively small.
Then equation \ref{eq:4} becomes
\begin{equation}
\frac{1}{\omega_{in}} = 
\frac{1}{\omega_o}
\sqrt{\frac{\signt({\rm NH_3})^2}{\signt({\rm N_2H^+})^2-\signt({\rm NH_3})^2}}~,
\end{equation}
and therefore be used to estimate the wavelength inside the coherent core as
\begin{equation}
\lambda_{in} = \sqrt{\frac{\pi}{\rho_o}}  
\frac{B_o}{\omega_o}
\sqrt{\frac{\signt({\rm NH_3})\, \signt({\rm N_2H^+})}{\signt({\rm N_2H^+})^2 - \signt({\rm NH_3})^2}}~. \label{eq:lambda_in}
\end{equation}

In the case of B5, we obtain a value of $\lambda_{in}\leq${\unit[0.64]{pc}}, which is larger than the diameter of the coherent core (\unit[0.34]{pc}). 
However, a direct estimation of the magnetic field strength using dust polarization measurements would substantially improve this estimate of the wavelength.

The critical length for wave propagation is obtained from equation 10.23 of \cite{StahlerPalla-2004} or equation 79 from \citealt{Mouschovias2011-MHD},
\begin{equation}
\lambda_{min} = \sqrt{\frac{\pi}{4\rho_o}} \frac{B_o}{\omega_o}~, \nonumber
\end{equation}
which we combine with eq.~\ref{eq:lambda_in} to write 
\begin{equation}
\frac{\lambda_{in}}{\lambda_{min}} = 2
\sqrt{\frac{\signt({\rm NH_3})\, \signt({\rm N_2H^+})}{\signt({\rm N_2H^+})^2 - \signt({\rm NH_3})^2}} = 3~.
\end{equation}
Therefore, the wavelength is sufficient to propagate the wave.

In addition, the characteristic damping timescale is given by equation 80 in \cite{Mouschovias2011-MHD}, 
\begin{equation}
    \tau_d = \frac{\lambda^2\,\omega_o}{2\pi^2\,v_A^2}~,
\end{equation}
where the Alfv\'en velocity is a function of density and magnetic field strength,
\begin{equation}
    v_A^2 = \frac{B_o^2}{4\pi\,\rho_o}~. \label{eq:Va}
\end{equation}
Replacing relations \ref{eq:lambda_in} and \ref{eq:Va} we obtain
\begin{equation}
    \tau_d = \frac{2}{\omega_o}\left(\frac{\signt({\rm NH_3})\, \signt({\rm N_2H^+})}{\signt({\rm N_2H^+})^2 - \signt({\rm NH_3})^2}\right) = 0.2~\textrm{Myr,}
\end{equation}
which is much smaller than the crossing time of the coherent region
\begin{equation}
    \tau_{cross} = 
    \frac{\rm diameter}{c_s} 
    = 1.8~\textrm{Myr,}
\end{equation}
and therefore the wave must be continually injected if they are to persist for a crossing time.

\subsection{Future observations}
Since it is unknown how does the amplitude of the penetrating wave damps inside the coherent dense core, it is important to take advantage of eq.~\ref{eq:deltaB} and observe different pairs of ion and neutral transitions with similar critical densities.
This would allow us to derive a damping of the penetrating wave in the subsonic region, which will be crucial to improve our understanding of this newly observed phenomenon.

One promising higher density tracer ion-neutral pair is N$_2$D$^+$ (1--0) and o-NH$_2$D (1$_{1,1}$--1$_{0,1}$).
These deuterated species are abundant in the densest regions of dense cores \citep[e.g.][]{Caselli2002_L1544-ions,Crapsi2007-L1544} and they selectively trace higher densities compared to the non-deuterated ones. 
It is also important that both these transitions have well known hyper-fine structure, which enables for a precise determination of the velocity dispersion.
Such comparison will provide a direct determination (or a strong upper limit) of the penetrating Alfv\'en wave damping.
These results would be a new direct observable constraint on the fragmentation process for numerical simulations including magnetic fields.

\section{Summary\label{sec-summary}}

We present new GBT Argus 8$\arcsec$ angular resolution observations of the Barnard 5 region in the \ntwohp (1--0) line and compare them with matched beam \ammo (1,1) and (2,2) VLA and GBT observations. 
Our results can be summarized as follows:
\begin{itemize}
\item Both tracers, \ntwohp and \ammo, show the presence of two filaments within the subsonic region (coherent core).
\item The centroid velocity maps for \ammo and \ntwohp display a similar pattern, with centroid velocities in  agreement within \unit[0.05]{\kms}.

\item The velocity dispersion maps show a narrow velocity dispersion throughout the region, but the \ntwohp line displays broader velocity dispersion than \ammo by about \unit[0.015]{\kms}.

\item The sonic Mach number for \ammo, 0.48, is smaller than for \ntwohp, 0.59, opposite to theoretical predictions of turbulent cascade or ambipolar diffusion for a static magnetic field.

\item The difference in sonic Mach number does not appear to be related to unresolved substructures, and it is fairly constant throughout the coherent core.

\item The observed difference between ions and neutrals is naturally explained by a penetrating Alfv\'en wave inside the coherent region, which affects more the ions than the neutrals. We estimate that the associated perturbation in the magnetic field is \unit[27]{$\mu$G}, which is at least 5\% of the previously estimated upper limit on the field in the region (\unit[500]{$\mu$G}, \citealt{Schmiedeke2021-B5}).

\item The wave's wavelength is 3$\times$ the critical wavelength, and it can propagate in the medium. Also, the damping timescale is $\approx 10$ times shorter than the sound wave crossing time of the coherent region, and therefore waves must be continually injected. 

\end{itemize}

\acknowledgments
JEP, PC, and AS acknowledge the support by the Max Planck Society. 
This material is based upon work supported by the Green Bank Observatory which is a major facility funded by the National Science Foundation operated by Associated Universities, Inc.
This research made use of APLpy, an open-source plotting package for Python.
This research made use of Astropy,\footnote{\url{http://www.astropy.org}} a community-developed core Python package for Astronomy. 
JEP thanks Stella Offner and Ralf Klessen for valuable discussions. We thank Paul Goldsmith, Stella Offner, Daniele Galli, Marco Padovani, Blakesley Burkhart, Shantanu Basu, and the anonymous referee for insightful comments that improved this paper.

\facility{GBT, VLA} 

\software{Aplpy \citep{aplpy,aplpy2019}, 
Astropy \citep{Robitaille_2013,Astropy2018}, 
Matplotlib \citep{Hunter_2007},
SciPy \citep{2020SciPy-NMeth},
pyspeckit \citep{pyspeckit},
spectral-cube \citep{spectral-cube}}

\appendix
\section{Sample spectra and best-model fit\label{Appendix-spectra}}

\begin{figure}[hb]
    \centering
    \includegraphics[width=0.45\textwidth]{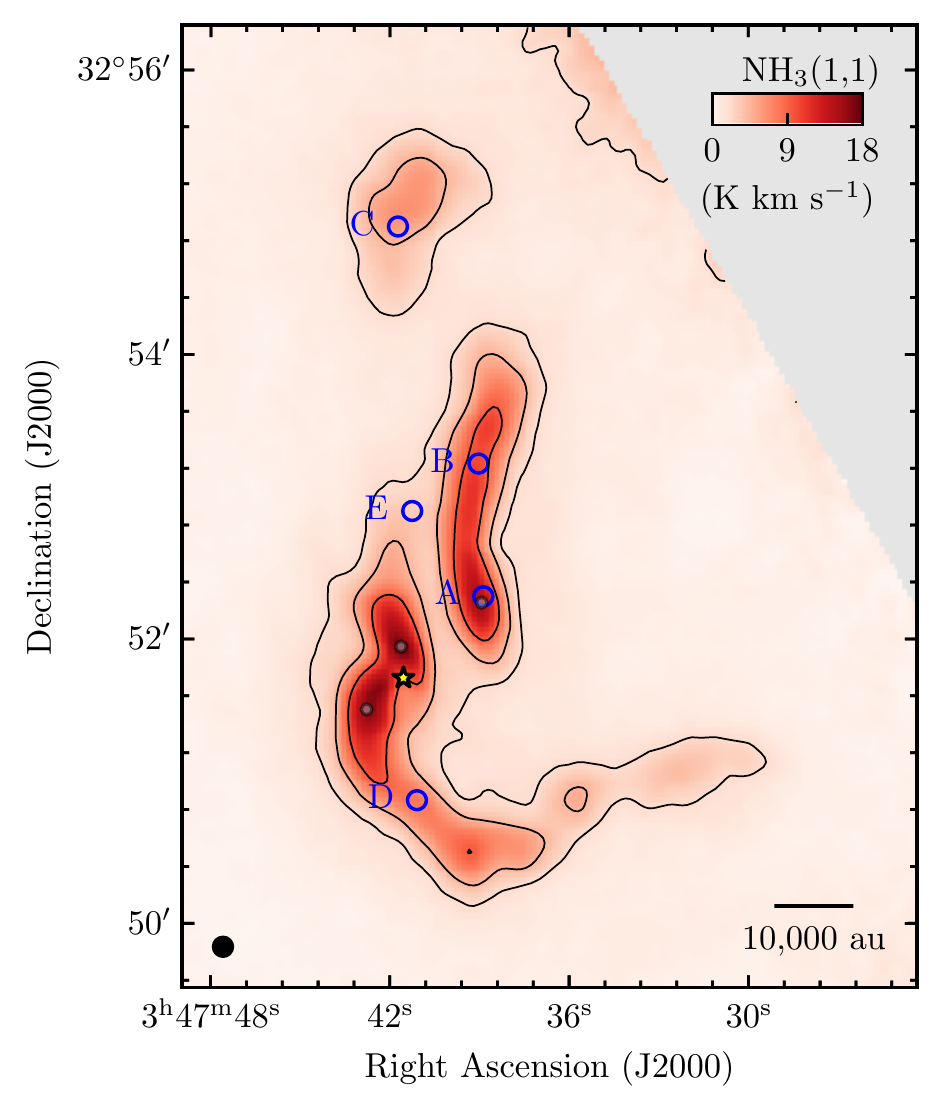}
    \includegraphics[width=0.54\textwidth]{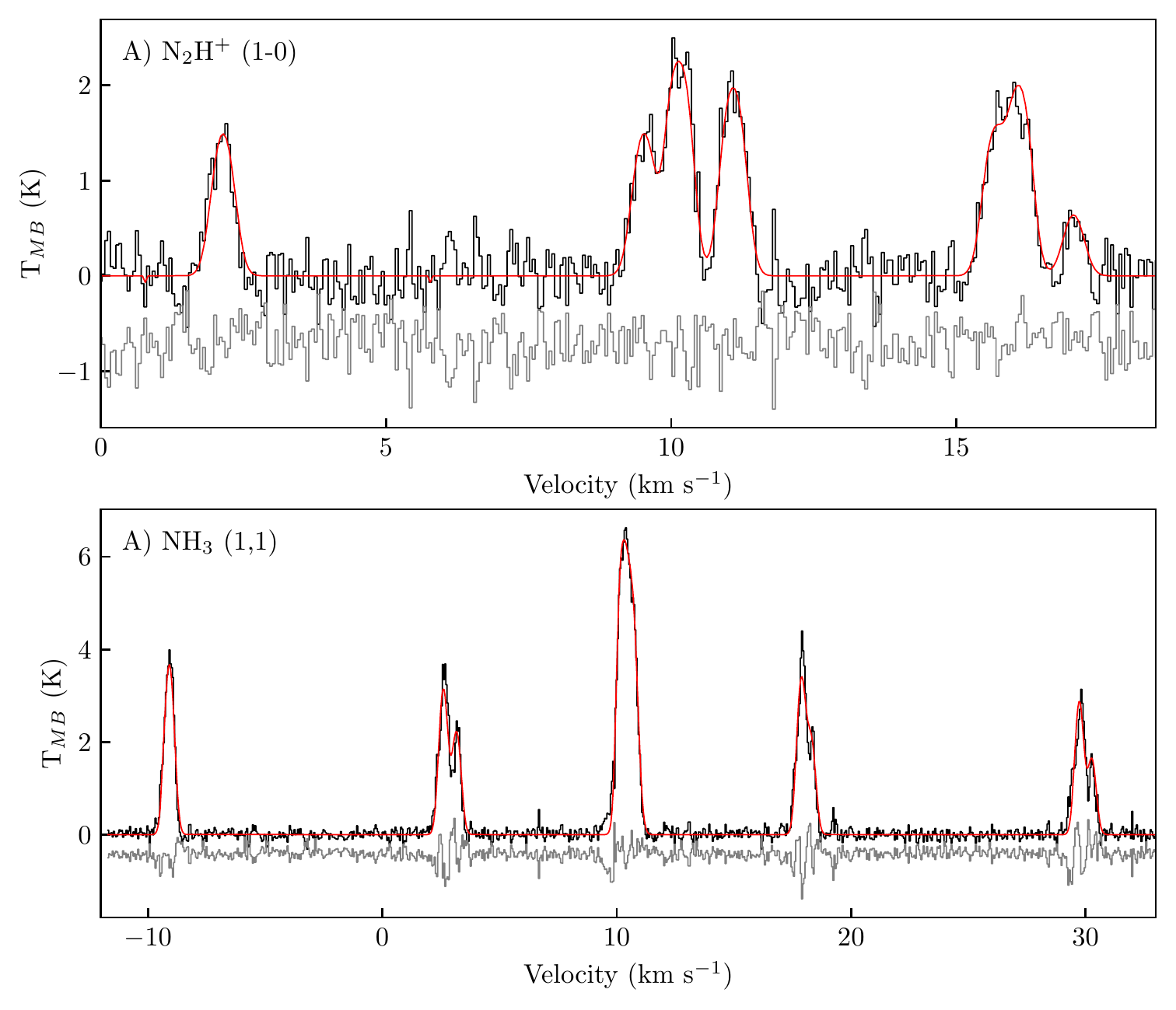}
    \caption{\emph{Left:} Integrated intensity map of \ammo(1,1), as in Figure~\ref{fig-TdV}. The blue circles show the positions of the sample spectra.
    \emph{Right:} Spectra obtained towards position A are shown. Top and bottom panels show the \ntwohp(1--0) and \ammo(1,1), respectively. 
    The data are shown in black, best fit model in red, and residual (model-data) is shown in gray (but offset for clarity).
    }
    \label{fig:app_map}
\end{figure}

\begin{figure}
    \centering
    \includegraphics[width=0.495\textwidth]{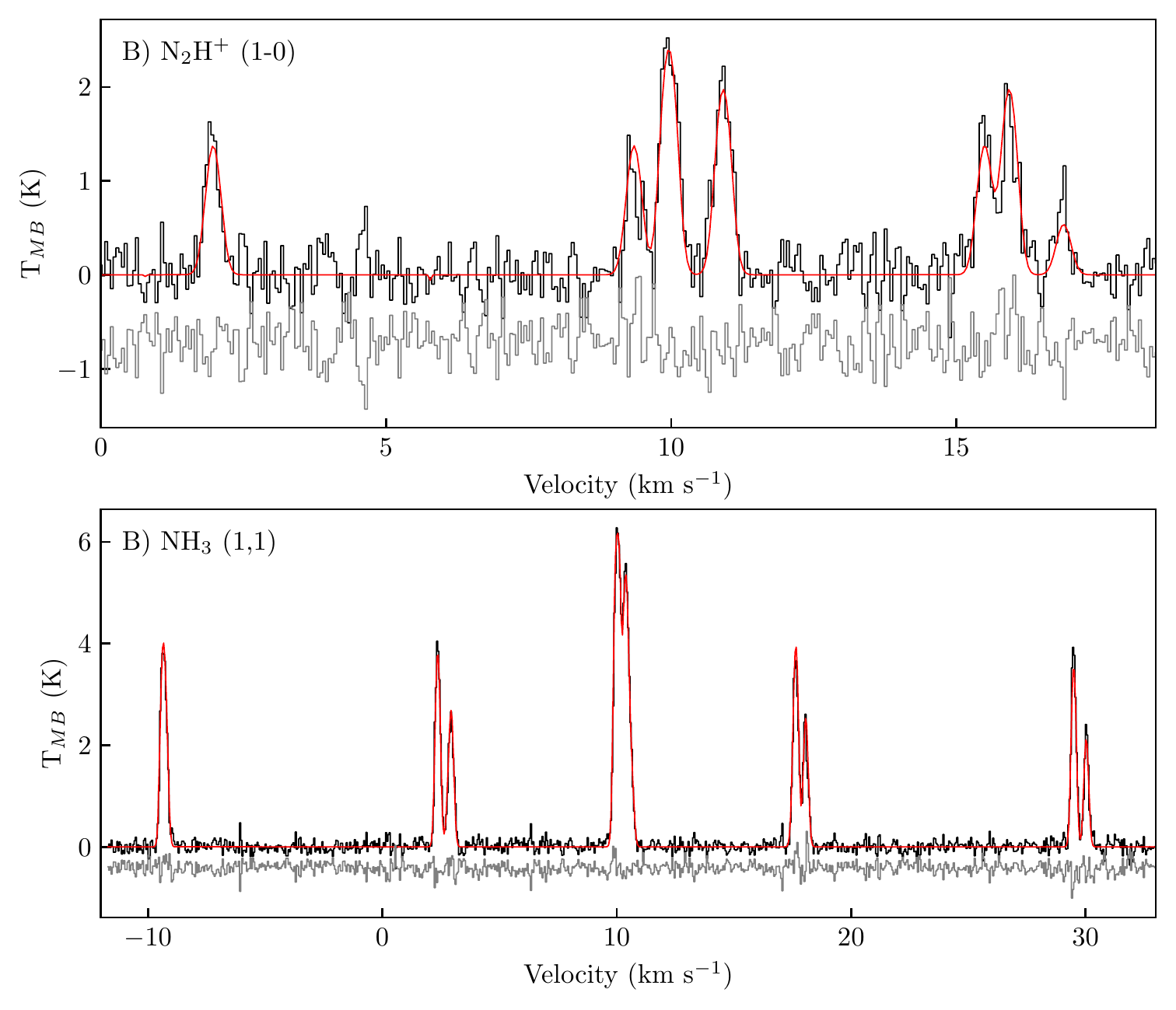}
    \includegraphics[width=0.495\textwidth]{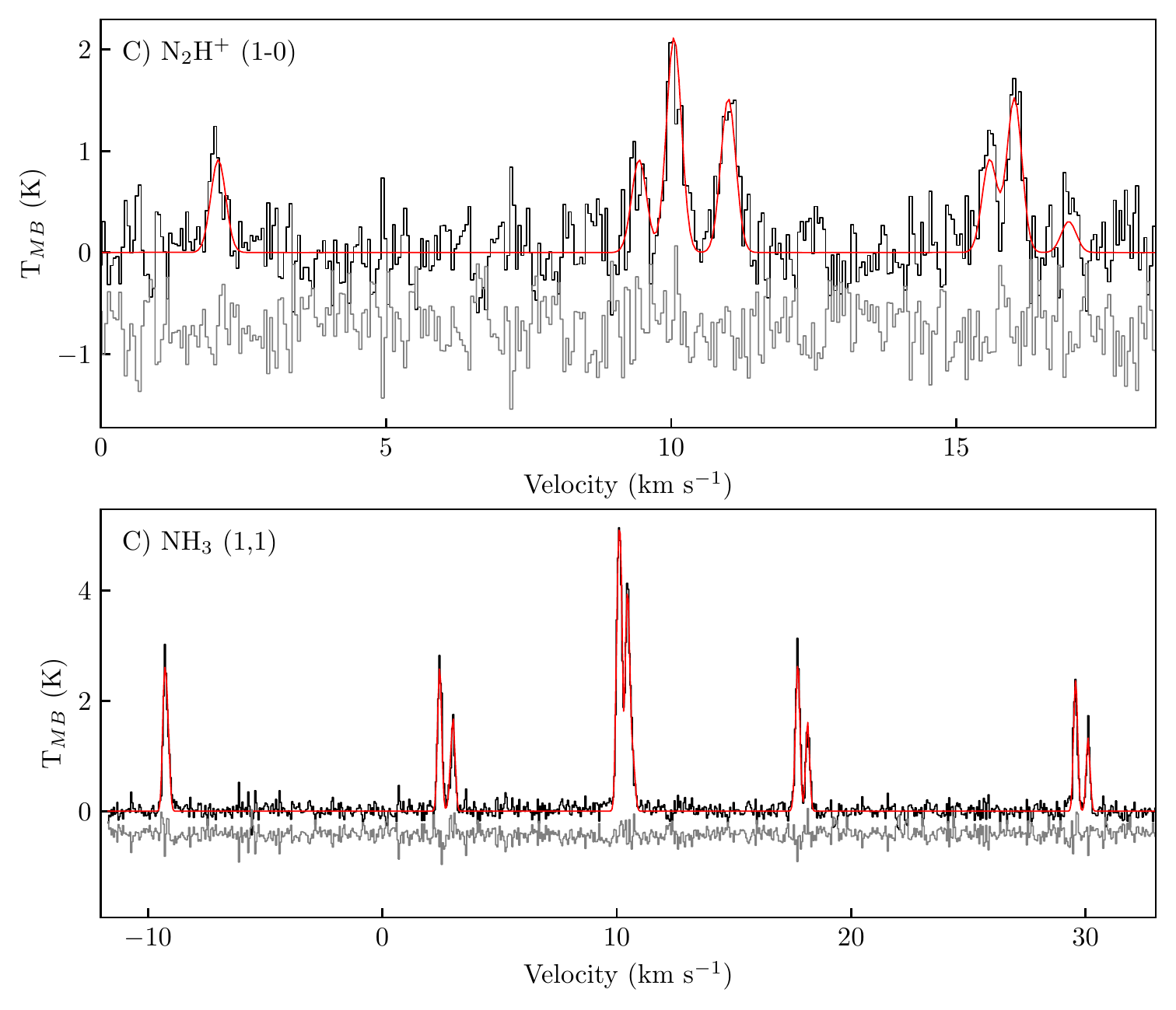}
    \caption{Spectra obtained towards positions B and C are shown in left and right panels, respectively. The top and bottom panels are as in Figure~\ref{fig:app_map}}
    \label{fig:app_map_BC}
\end{figure}

\begin{figure}
    \centering
    \includegraphics[width=0.495\textwidth]{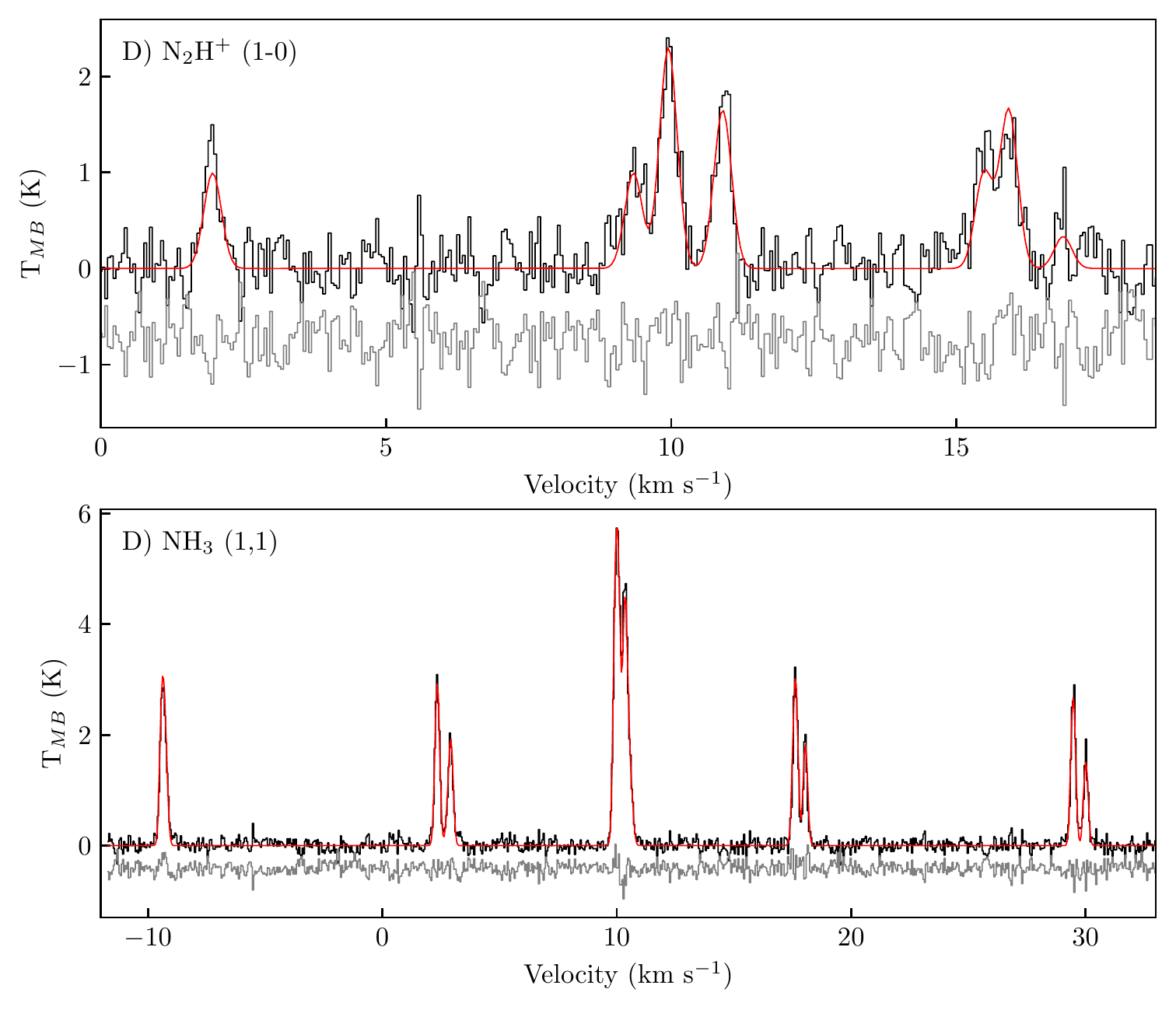}
    \includegraphics[width=0.495\textwidth]{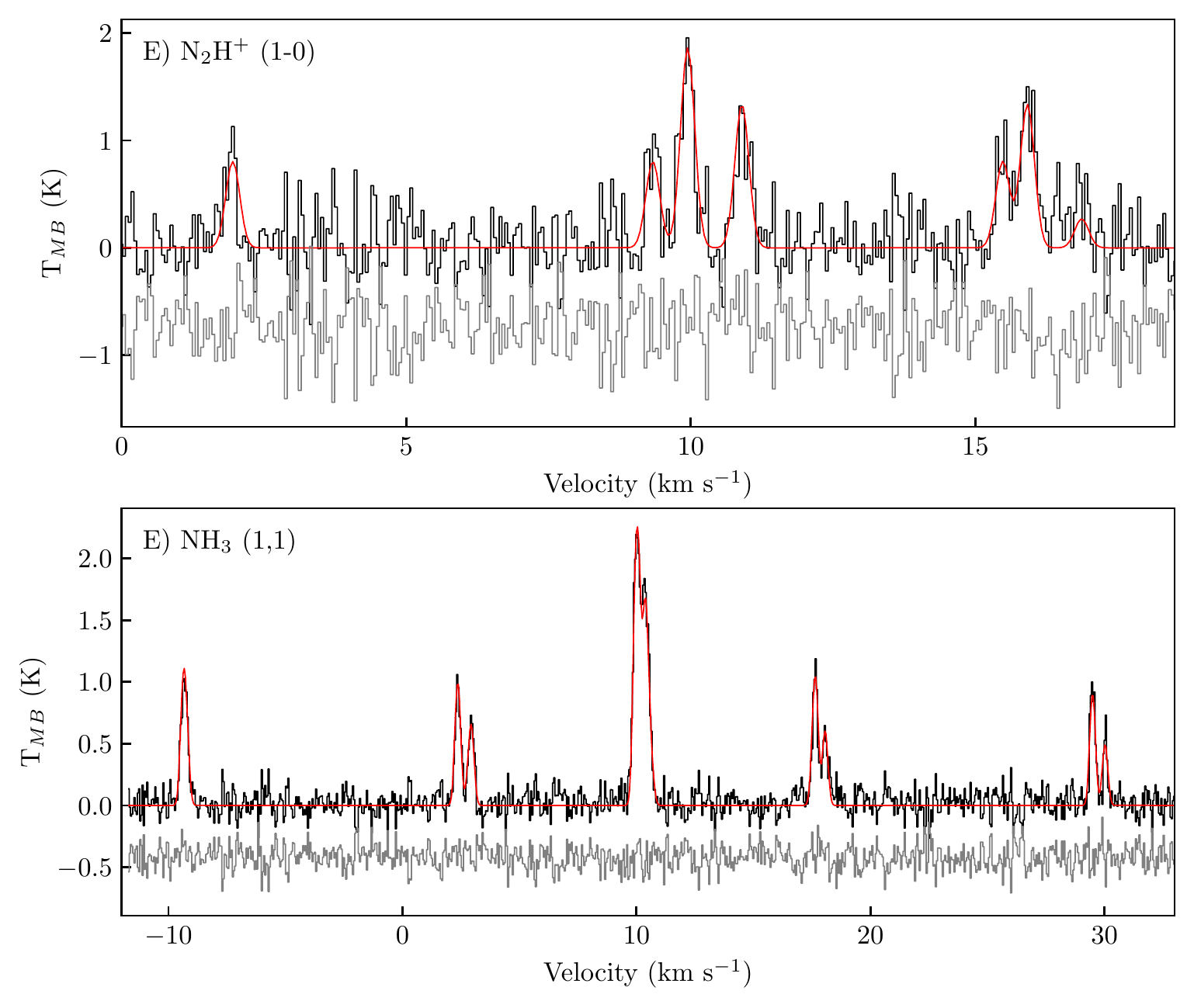}
    \caption{Spectra obtained towards positions D and E are shown in left and right panels, respectively. The top and bottom panels are as in Figure~\ref{fig:app_map}}
    \label{fig:app_map_DE}
\end{figure}

We show the \ntwohp (1--0) and \ammo(1,1) spectra towards a sample of representative regions in the map. 
The left panel of Figure~\ref{fig:app_map} shows the \ammo(1,1) integrated intensity map (as in Figure~\ref{fig-TdV}) with blue circles marking the positions for the shown spectra.
The positions selected include: 
A) peak of B5-Cond1, 
B) high column density in the narrow filament,
C) isolated dense core at the north, 
D) lower column density in the filament, and 
E) low column density in outside the filaments but inside the coherent zone.
The \ntwohp and \ammo spectra toward position A are shown in Figure~\ref{fig:app_map}, 
positions B and C are shown in Figure~\ref{fig:app_map_BC}, and 
positions D and E are shown in Figure~\ref{fig:app_map_DE}.
The data in these figures are shown in black, while the best-fit models are shown in red, and the residuals (model - data) are shown offset and in gray.
In general, all these figures show that the best-models provide a good fit to the data.

\bibliographystyle{aasjournal}

\bibliography{bibliography}

\end{document}